\documentclass[aps,prl,showpacs,twocolumn,superscriptaddress,floatfix]{revtex4}
\usepackage{graphicx}
\usepackage{amsmath,amssymb}
\newcommand{\un}[1]{\,\mathrm{#1}}
\usepackage{color}

\begin{document}

\title{Phonon-mediated vs.\ Coulombic Back-Action in Quantum Dot circuits}

\author{D.\ Harbusch}

\affiliation{Center for NanoScience and Fakult\"at f\"ur Physik, Ludwig-Maximilians-Universit\"at M\"unchen, Geschwister-Scholl-Platz 1, D-80539 M\"unchen, Germany}

\author{D.\ Taubert}

\affiliation{Center for NanoScience and Fakult\"at f\"ur Physik, Ludwig-Maximilians-Universit\"at M\"unchen, Geschwister-Scholl-Platz 1, D-80539 M\"unchen, Germany}
\author{H.\,P.\ Tranitz}

\affiliation{Institut f\"ur Experimentelle Physik, Universit\"at Regensburg, D-93040 Regensburg, Germany}

\author{W.\ Wegscheider}

\affiliation{Laboratory for Solid State Physics, ETH Z\"urich, CH-8093 Z\"urich, Switzerland}

\author{S.\ Ludwig}

\affiliation{Center for NanoScience and Fakult\"at f\"ur Physik, Ludwig-Maximilians-Universit\"at M\"unchen, Geschwister-Scholl-Platz 1, D-80539 M\"unchen, Germany}

\begin{abstract}
Quantum point contacts (QPCs) are commonly employed to detect capacitively the charge state of coupled quantum dots (QD). An indirect back-action of a biased QPC onto a double QD laterally defined in a GaAs/AlGaAs heterostructure is observed. Energy is emitted by non-equilibrium charge carriers in the leads of the biased QPC. Part of this energy is absorbed by the double QD where it causes charge fluctuations that can be observed under certain conditions in its stability diagram. By investigating the spectrum of the absorbed energy, we identify both acoustic phonons and Coulomb interaction being involved in the back-action, depending on the geometry and coupling constants.
\end{abstract}

\pacs{03.67.-a, 63.22.-m, 72.70.+m, 73.21.La, 73.23.Hk}

\maketitle
Coupled quantum dots (QDs) are promising candidates for applications as qubits in solid state quantum information processing schemes \cite{Loss1998}. One important criterion is the scalability of the qubit number. In a complex layout it will pose a great challenge to implement readout techniques that address singe qubits without adding decoherence to the coupled QDs. Direct transport through an array of QDs is limited due to Coulomb blockade. However, a single biased quantum point contact (QPC) in a separate circuit can act as charge detector for several QDs \cite{Field1993,Elzerman2003}. QPCs are straightforward to implement, yield sufficient sensitivity, and can be operated as wide bandwidth detectors \cite{Reilly2007,Cassidy2007}. The latter is desirable in quantum information processing where a rapid detection scheme is needed. The suitability of QPCs as fast detectors has been demonstrated in single-shot readout \cite{Elzerman2003-2,barthel2009} and counting statistics experiments \cite{Gustavsson2006,Fricke2007}. Increasing the bandwidth, however, requires a high signal-to-noise ratio which makes it necessary to operate the QPC at a relatively high bias voltage.

A biased QPC employed as a charge detector causes back-action. Its quantum limit can be traced back to statistical charge fluctuations at the QPC capacitively coupled to QDs \cite{Young2009}. Shot noise only contributes to back-action if the QPC has resistive leads \cite{Young2009}. In addition to these direct Coulomb back-action mechanisms, the solid state environment provides possibilities for indirect back-action \cite{Khrapai2006,Khrapai2007,Taubert2008,Gasser2009}. A biased QPC emits non-equilibrium charge carriers into its leads that then relax via electron-electron interaction, the emission of plasmons, or acoustic phonons \cite{Ridley1991}. Partial reabsorption of the emitted energy can result in charge fluctuations in (coupled) QDs, hence causing indirect back-action. Usually these fluctuations are too fast to be detected in measurements with limited bandwidth, but under certain conditions they can be observed in the stability diagram of coupled QDs \cite{Taubert2008}. In this Letter we present a systematic investigation of such back-action-induced charge fluctuations in a double QD. We find that both acoustic phonons and Coulomb interaction can play an important role for the back-action in realistic devices.

\begin{figure}
\includegraphics{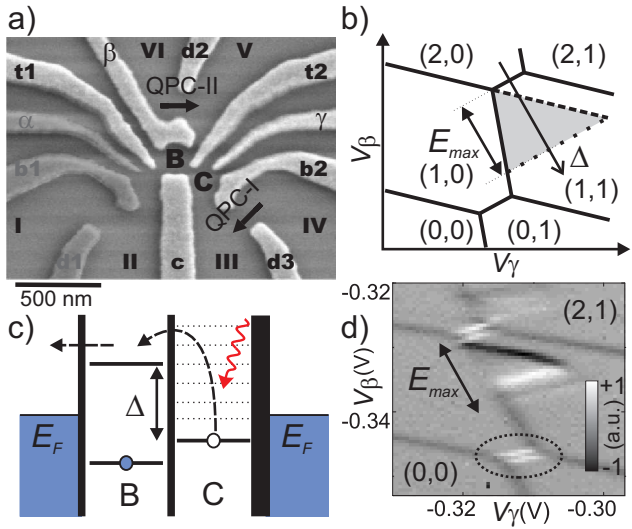}
\caption{(a)~Scanning electron micrograph of a nominally identical device. Metal gates (light gray) are negatively biased, darker gates are grounded. QDs B and C, current paths (arrows), and ohmic contacts (roman numbers) are indicated. (b) Sketch of a double QD charge stability diagram. Numbers in brackets indicate stable charge configurations ($N_B$, $N_C$) of the double QD. The gray triangle features back-action (compare d). (c)~Level diagram of the double QD. Dotted lines depict the electron excitation spectrum of QD C. (d)~Measured transconductance $dI_{\mathrm{QPC}}/dV_{\beta}$ (gray scale) as a function of voltages applied to gates $\beta$ and $\gamma$ for $V_{\mathrm{QPC}} = -1.6\,$mV and $P_{\mathrm{QPC}}= 0.64\,$pW.}
\label{fig1}
\end{figure}

Our device is based on a GaAs/AlGaAs heterostructure containing a two-dimensional electron system 90\,nm beneath the surface. Charge carrier density and mobility are $n_{e} = 2.78\times 10^{15}\mathrm{\un{m^{-2}}}$ and $\mu = 140\mathrm{\un{m^{2}/Vs}}$. QDs and QPCs are electrostatically defined by applying negative voltages to metallic gates fabricated by e-beam lithography. The gate layout is shown in Fig.\ \ref{fig1}a.
The measurements are performed at an electron temperature of $T_\mathrm{el} \lesssim 130\un{mK}$. Although the gate layout is designed for three QDs \cite{Schroer2007}, here we define only a double QD (gates {\it d1, b1} and $\alpha$ are grounded). Unless otherwise stated, only one of the implemented QPCs (black arrows in Fig.\ \ref{fig1}a), namely QPC-I, is biased by applying a voltage $V_{\mathrm {QPC}}$ to contact IV. Since all other contacts are grounded and QPC-I is operated near pinch-off, the double QD is virtually unbiased. The dc current $I_{\mathrm{QPC}}$ flowing through QPC-I is measured with a bandwidth of only 10\,Hz; $I_{\mathrm{QPC}}$ therefore probes the \emph{average} charge configuration of the double QD. We obtain transconductance data $dI_{\mathrm{QPC}}/dV_{\beta}$ by numerical differentiation of $I_{\mathrm{QPC}}$. We have observed back-action in a wide range of charge configurations, but here we focus on two electrons or less occupying the double QD. Solid lines in Fig.\ \ref{fig1}b sketch the expected charge stability diagram. Ground state configurations are denoted $(N_{B},N_{C})$, indicating that QD B (C) is occupied by $N_B$ ($N_C$) electrons.

For the experiments presented here, it is essential to adjust the tunnel couplings of the double QD to be very asymmetric \cite{Taubert2008}. In a symmetric configuration, fundamental laws of thermodynamics prevent the observation of the back-action effects discussed here (see supplementary material \cite{suppl}). In our case the right tunnel barrier $b2$ between QD C and lead III is almost closed, resulting in a tunneling rate of only $\Gamma_{\mathrm{b2}} \simeq 25$\,kHz. The interdot tunneling rate $\Gamma_{\mathrm{t2}} \simeq 0.7$ GHz between QD B and C, as well as $\Gamma_{\mathrm{t1}}$ between QD B and lead II are much higher. The measured charge stability diagram in Fig.\ \ref{fig1}d shows the transconductance $dI_{\mathrm{QPC}}/dV_{\beta}$ in gray scale as a function of the gate voltages $V_{\beta}$ and $V_{\gamma}$. Two deviations from the usual honeycomb pattern (without signs of back-action) are observed. First, charge reconfiguration lines are split into double lines (circled in Fig.\ \ref{fig1}d). In between these two white lines the dc current $I_{\mathrm{QPC}}$ versus $V_{\beta}$ exhibits a plateau at a value reflecting an equal occupation of the configurations (1,0) and (0,1) \cite{suppl}. This can be explained with rapid transitions between the symmetric and antisymmetric combinations of the two almost degenerate configurations \cite{Rushforth2004}. The energy source driving these transitions is discussed below.

The second irregularity is a triangular-shaped region in the center of Fig.\ \ref{fig1}d described in more detail in Ref.\ \cite{Taubert2008}. Within the triangle, the charge in one of the QDs (here QD C) fluctuates. An electron from QD C tunnels to QD B and from there into lead II and vice versa: (1,1) $\leftrightarrow$ (2,0) $\leftrightarrow$ (1,0). As can be seen in Fig.\ \ref{fig1}c, which shows the chemical potentials of the QDs and leads, these charge fluctuations require the absorption of energy. One of the border lines of the triangle in Fig.\ \ref{fig1}d is parallel to the charge reconfiguration lines (white double lines). Along this border line the energy difference $\Delta$ (asymmetry energy) between the ground state configurations (1,1) and (2,0) is thus constant (Figs.\ \ref{fig1}b and \ref{fig1}c). With the charging energy of QD B ($2.5\,$meV) the size of the triangle can be converted into an energy $E_\mathrm{max}$ (Figs.\ \ref{fig1}b and \ref{fig1}d). We interpret $E_\mathrm{max}$ as the maximum energy that QD C absorbs in a single process.

\begin{figure}
\includegraphics{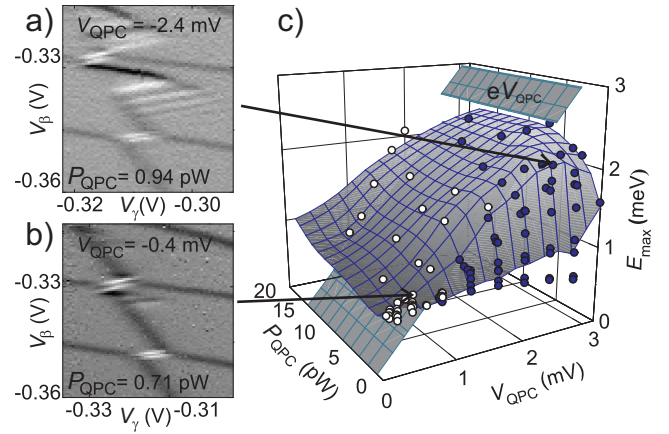}
\caption{(a,b): Transconductance $dI_{\mathrm{QPC}}/dV_{\beta}$ (gray scale) as a function of $V_{\beta}$ and $V_{\gamma}$. (c) Observed maximum of absorbed energy $E_{\mathrm{max}}$ (compare Fig.\ \ref{fig1}d) as a function of $P_{\mathrm{QPC}}$ and $V_{\mathrm{QPC}}$. Arrows show where (a) and (b) are located in this graph. See main text for details.}
\label{fig2}
\end{figure}

Figs.\ \ref{fig2}a and \ref{fig2}b
plot stability diagrams similar to that in Fig.\ \ref{fig1}d for two very different bias voltages $V_{\mathrm{QPC}}$. The triangle size clearly grows with increasing bias indicating that QPC-I acts as energy source. Fig.\ \ref{fig2}c underlines this result. $E_{\mathrm{max}}$ is plotted as a function of $V_{\mathrm{QPC}}$ and the dissipated power $P_{\mathrm{QPC}}=I_{\mathrm{QPC}}V_{\mathrm{QPC}}$
 with each data point corresponding to the size of one triangle. The curved surface fitted to the data is a guide to the eye. The gray plane in Fig.\ \ref{fig2}c is defined by $E_{\mathrm{max}}=e\,V_{\mathrm{QPC}}$. This is the largest energy quantum the QPC can emit and the expected $E_{\mathrm{max}}$ for back-action mediated by direct (first-order) Coulomb interaction, as compellingly suggested by previous data for the case of shot-noise \cite{Gustavsson2007}. In our measurements, the open circles lie above this plane while the closed circles are below it. The apparently missing clear cut-off at $E_{\mathrm{max}}=e\,V_{\mathrm{QPC}}$ \cite{Gustavsson2007} in Fig.\ \ref{fig2}c suggests that in our data direct back-action is unimportant. This has to be seen in the context of the very small conductance of our QPC, $G_{\mathrm{QPC}}\ll 0.5\, G_{\mathrm{0}}$, where $G_{\mathrm{0}}=2e^{2}/h$. In this regime the direct back-action related to shot noise of $I_{\mathrm{QPC}}$ or charge fluctuations at the QPC is expected to be strongly suppressed \cite{Young2009,Reznikov1995,suppl}. Especially, the deviations at large $V_{\mathrm{QPC}}$, where we find $E_\mathrm{max}<eV_{\mathrm{QPC}}$, imply indirect back-action with energy dissipated in the leads of the QPC. During the relaxation processes the energy spectrum is likely shifted towards lower energies before some of the resulting energy quanta are reabsorbed by the double QD.

$E_{\mathrm{max}}$ strongly increases with $P_{\mathrm{QPC}}$ before it saturates at $P_{\mathrm{QPC}}\simeq 0.5\,$pW (Fig.\ \ref{fig2}c). The observation $E_{\mathrm{max}}<eV_\mathrm{QPC}$ for very small $P_{\mathrm{QPC}}$ is another indication that the high energy end of the spectrum emitted by the QPC is suppressed in the absorption spectrum of the double QD. This again suggests indirect back-action where multiple scattering processes in the leads of the QPC alter the original emission spectrum.

Our scenario of indirect back-action fails to explain $E_{\mathrm{max}}>e\,V_{\mathrm{QPC}}$ in the regime of small $V_{\mathrm{QPC}}$, and is in contrast to the observed lower bound of $E_\mathrm{max}\simeq0.6\,$meV (open circles in Fig.\ \ref{fig2}c). In the present experiment, this marks the limit of external non-thermal noise. It mainly consists of 50\,Hz signals which stem from electronic instruments. Apparently, an asymmetrically coupled double QD can be employed as a sensitive noise detector.

The stability diagrams
\begin{figure}
\includegraphics{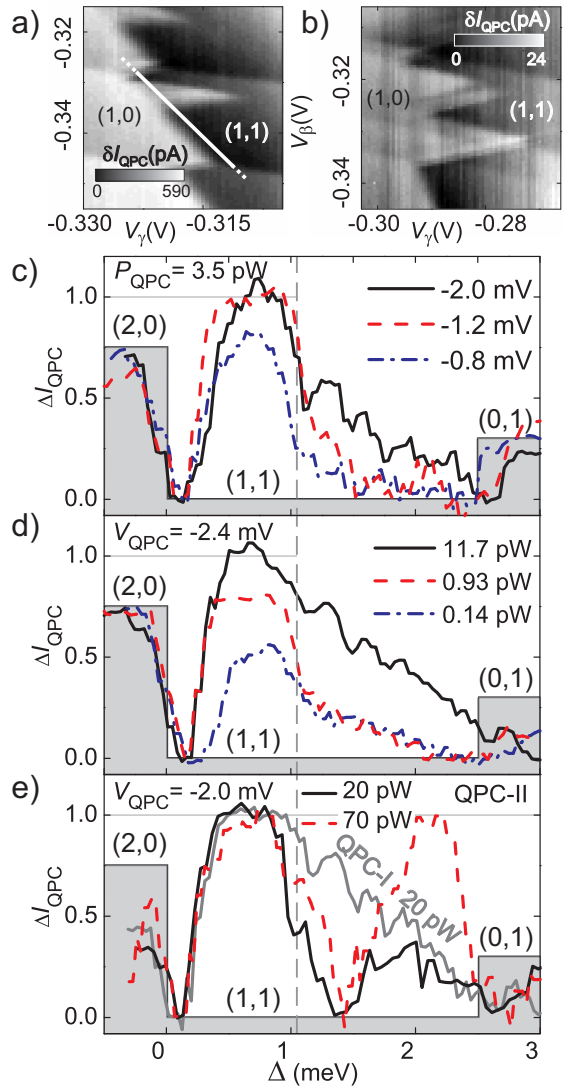}
\caption{(a,b) $\delta I_{\mathrm{QPC}}$ (gray scale) as a function of $V_{\beta}$ and $V_{\gamma}$. A plane fit is subtracted from $I_{\mathrm{QPC}}$ (resulting in $\delta I_{\mathrm{QPC}}$) to correct for capacitances between gates and the QPC. In (a) only QPC-I is biased with $V_{\mathrm{QPC}}=-0.8\,$mV and $P_{\mathrm{QPC}}=2.5\,$pW; for (b) the values are $V_{\mathrm{QPC}}=-0.1\,$mV and $P_{\mathrm{QPC}}=0.01\,$pW. QPC-II is additionally biased in with $V_{\mathrm{QPC}}=-2.0\,$mV and $P_{\mathrm{QPC}}=72\,$pW in (b). (c--e) Normalized current change $\Delta I_{\mathrm{QPC}}$ versus asymmetry energy $\Delta$ (compare Fig.\ \ref{fig1}c) along the white line in (a). $\Delta I_{\mathrm{QPC}}$ corresponding to configurations (2,0), (1,1) and (0,1) are highlighted in gray.}
\label{fig3}
\end{figure}
in Figs.\ \ref{fig3}a and \ref{fig3}b plot the current change $\delta I_{\mathrm{QPC}}$ at QPC-I (also compare Fig.\ \ref{fig1}b). In Fig.\ \ref{fig3}a back-action is visible in the shape of a triangle of enhanced $\delta I_{\mathrm{QPC}}$. Converting the gate voltage along the white line in Fig.\ \ref{fig3}a into the asymmetry energy $\Delta$ (defined in Fig.\ \ref{fig1}c) we plot cross sections through such triangles in Figs.\ \ref{fig3}c -- \ref{fig3}e. In thermal equilibrium we expect the double QD to occupy its ground state configuration because of the low $T_\mathrm{el}\simeq 130\,$mK. The y-axis shows $\Delta I_{\mathrm{QPC}}$ which is calculated from $I_{\mathrm{QPC}}$ by subtracting the equilibrium value in configuration (1,1). To achieve comparability, the curves are scaled so that $\Delta I_{\mathrm{QPC}}=1$ in (1,0). The white line in Fig.\ \ref{fig3}a starts in configuration (2,0), crosses the area of (1,1) and ends in (0,1). The corresponding average values of $\Delta I_{\mathrm{QPC}}$ measured at these configurations are indicated in Figs.\ \ref{fig3}c -- \ref{fig3}e as a shaded background. Fig.\ \ref{fig3}c displays curves for different $V_\mathrm{QPC}$ in the high power limit of Fig.\ \ref{fig2}c, while the power dependence is investigated in Fig.\ \ref{fig3}d. All these curves follow approximately the behavior expected for thermal equilibrium (shaded background). Deviations are observed only at the back-action induced triangles located in the (1,1) area. Within these triangles the data display a general trend of an increasing $\Delta I_{\mathrm{QPC}}$ with growing $V_\mathrm{QPC}$ and $P_\mathrm{QPC}$. However, $\Delta I_{\mathrm{QPC}}\simeq1$ represents an upper limit for all our measurements. Since the intermediate configuration (2,0) decays very fast into (1,0), $\Delta I_{\mathrm{QPC}}$ indicates the average occupation number difference of the configurations (1,1) and (1,0). Wherever $\Delta I_{\mathrm{QPC}}\simeq1$ in Figs.\ \ref{fig3}c -- \ref{fig3}e, the higher energy configuration (1,0) is strongly occupied. This can only be explained in terms of a non-equilibrium energy source driving the transitions.

All triangles induced by back-action of QPC-I can be divided into two regimes. For $0<\Delta\lesssim1.04\,$meV (vertical dashed line in Figs.\ \ref{fig3}c -- \ref{fig3}e) we observe a featureless region where the current tends to saturate at $\Delta I_{\mathrm{QPC}}\simeq1$. At $\Delta\simeq1.04\,$meV the current sharply drops, and for $\Delta>1.04\,$meV, we find $\Delta I_{\mathrm{QPC}}<1$ even for large $V_\mathrm{QPC}$ and $P_\mathrm{QPC}$. This region, however, features an additional substructure, best seen in Fig.\ \ref{fig2}a, namely lines of constant transconductance parallel to the charge reconfiguration lines. These lines correspond to constant detuning $\Delta$ between (1,1) and (2,0). They reveal the quantum mechanical excitation spectrum of QD C (compare Fig.\ \ref{fig1}c) \cite{comment-1}. Whenever an electron in QD C is lifted to an excited state (1,1)*, that is in resonance with the ground state of (2,0), tunneling between the two QDs is enhanced. This leads to the observed alternating occupation probability as a function of $\Delta$. Although we expect additional excited states of QD C, we do not observe them at $\Delta<1.04\,$meV.

We explain the sharp current drop at $\Delta\simeq1.04\,$meV as follows. In Ref.\ \cite{Schinner2009} phonon-mediated interaction between mesoscopic circuits has been demonstrated. Back-scattering of an electron defines an upper limit $E_\mathrm{max}^\mathrm{ph} \simeq 2\hbar k_{\mathrm{F}} v_{\mathrm{s}}$ for the energy that can be transferred to an acoustic phonon \cite{Schinner2009}. With our Fermi energy of $E_\mathrm F\simeq 10\,$meV and the maximum sound velocity $v_\mathrm s\simeq6000\,$m$/$s from Ref.\ \cite{Schinner2009}, we find $E_\mathrm{max}^\mathrm{ph}\simeq1.04\,$meV. Just at this asymmetry $\Delta=E_\mathrm{max}^\mathrm{ph}$ the current drops sharply (Figs.\ \ref{fig3}c--e). We conclude that for $\Delta\lesssim1.04\,$meV the back-action is mainly caused by phonons emitted in the leads of the biased QPC and reabsorbed by a QD. In principle, absorption of multiple phonons could account for back-action observed for $\Delta>1.04\,$meV. However, the existence of two different interaction mechanisms seems more likely, because of the observation of the excitation spectrum of QD C only for $\Delta>1.04\,$meV.

For the data shown in Figs.\ \ref{fig3}b and \ref{fig3}e, QPC-II is strongly biased and used as energy emitter (while the weakly biased QPC-I is still the detector). Fig.\ \ref{fig3}e displays two measurements for QPC-II as emitter and one measurement for QPC-I (gray solid line) as emitter. When QPC-II is strongly driven, $\Delta I_\mathrm{QPC}$ drops all the way to zero near $\Delta\simeq1.04\,$meV. Then, for $\Delta>1.04\,$meV, a second triangle of charge fluctuations appears, as can be best observed in Fig.\ \ref{fig3}b. Both triangles have no substructure. This is in direct contrast to the results obtained with QPC-I as emitter, where we observe a characteristic substructure for $\Delta>1.04\,$meV (parallel lines in Fig.\ \ref{fig2}a), namely the excitation spectrum of QD C.

An important difference between the two QPCs is, that the capacitive coupling between the double QD and QPC-I is roughly twice as large compared to QPC-II. Experimentally we find that the excitation spectrum of QD C can only be resolved if QPC-I is emitter, where the capacitive coupling between QD C and the energy emitting QPC (and its leads) is strong. These results imply that Coulomb interaction is the dominant back-action mechanism for $\Delta>1.04\,$meV. At the same time, the observed back-action must be indirect (as discussed above, see Fig.\ \ref{fig2}c). We suggest a mechanism in which non-equilibrium charge carriers are emitted by QPC-I to lead III. Next, excited carriers in lead III exchange energy with QD C via Coulomb interaction. This scenario explains the remaining back-action for $\Delta>1.04\,$meV in the case of QPC-I being emitter.

The position of the second (lower) triangle in Fig.\ \ref{fig3}b indicates transitions involving the configurations (1,1)\,$\leftrightarrow$\,(0,1)\,$\leftrightarrow$\,(1,0) compared to (1,1)\,$\leftrightarrow$\,(2,0)\,$\leftrightarrow$\,(1,0) for the upper triangle. Here the electron in QD B tunnels to lead II after absorbing energy, then the electron in QD C relaxes to QD B (and emits energy). To first order, the transition (1,1)\,$\rightarrow$\,(0,1) cannot be driven by phonons ((1,1)\,$\rightarrow$\,(2,0) for the upper triangle) whenever the energy difference between these two configurations exceeds $E_\mathrm{max}^\mathrm{ph}\simeq1.04\,$meV. In fact, the size of both triangles in Fig.\ \ref{fig3}b (and the width of both local maxima in Fig.\ \ref{fig3}e) are identical and equal to $E_\mathrm{max}^\mathrm{ph}$. This result strongly points to phonon-mediated back-action for both triangles. Note that for $P_{\mathrm{QPC}} >15\,$pW, the second phonon-mediated triangle is also weakly visible with QPC-I as emitter \cite{suppl}. The apparent difference in interaction strength can be partly explained by the anisotropic coupling tensors between electrons and phonons and the sample geometry.

In conclusion, we demonstrate a method to directly measure back-action of a biased QPC on a double QD causing charge fluctuations. Back-action spectroscopy allows us to identify phonon-induced back-action as well as features most likely caused by Coulomb interaction. The observed back-action is indirect in nature, distinguishing it from the direct Coulomb interaction between charge fluctuations at the QPC and the electrons confined in QDs. Comparing two different QPCs reveals a strong dependence of the back-action on geometry which needs further investigation. Our results will help to develop detectors with reduced back-action.

We thank J.\,P.\ Kotthaus, G.\,J.\ Schinner, T.\ Ihn, and D.\,M.\ Eigler for fruitful discussions. Financial support by the German Science Foundation via SFB 631 and SFB 689, the German Israel program DIP, the German Excellence Initiative via the "Nanosystems Initiative Munich (NIM)", and LMUinnovativ (FuNS) is gratefully acknowledged.




\end{document}


\title{Phonon-mediated vs.\ Coulombic Back-Action in Quantum Dot circuits:\\ Electronic Supplementary}

\author{D.\ Harbusch}

\affiliation{Center for NanoScience and Fakult\"at f\"ur Physik, Ludwig-Maximilians-Universit\"at M\"unchen, Geschwister-Scholl-Platz 1, D-80539 M\"unchen, Germany}

\author{D.\ Taubert}

\affiliation{Center for NanoScience and Fakult\"at f\"ur Physik, Ludwig-Maximilians-Universit\"at M\"unchen, Geschwister-Scholl-Platz 1, D-80539 M\"unchen, Germany}
\author{H.\,P.\ Tranitz}

\affiliation{Institut f\"ur Experimentelle Physik, Universit\"at Regensburg, D-93040 Regensburg, Germany}

\author{W.\ Wegscheider}

\affiliation{Laboratory for Solid State Physics, ETH Z\"urich, CH-8093 Z\"urich, Switzerland}

\author{S.\ Ludwig}

\affiliation{Center for NanoScience and Fakult\"at f\"ur Physik, Ludwig-Maximilians-Universit\"at M\"unchen, Geschwister-Scholl-Platz 1, D-80539 M\"unchen, Germany}

\maketitle


\section{Characterization and working point of QPC-I}
\begin{figure}[h]
\includegraphics[width=16cm]{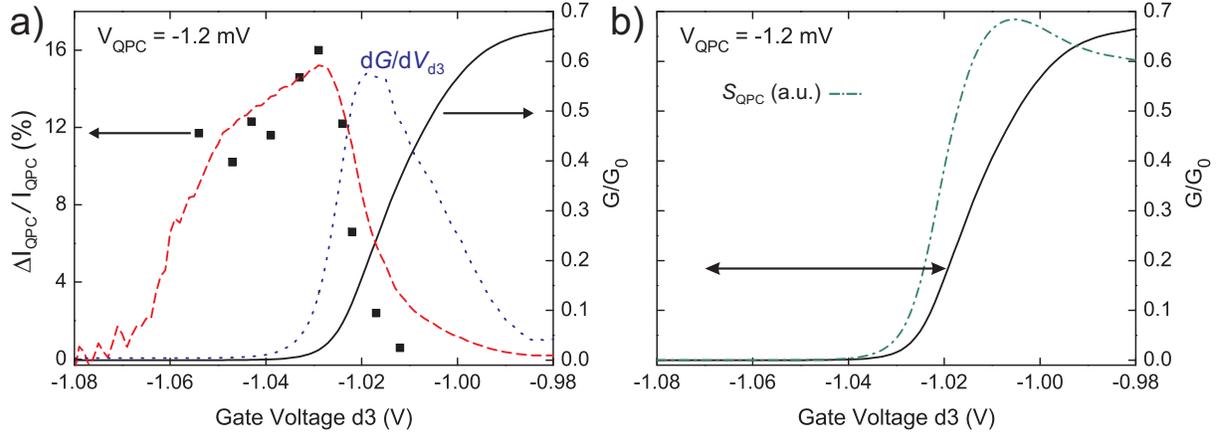}
\caption{The black line in a and b plots the conductance (axis on the right) of QPC-I as a function of gate voltage $V_\mathrm{d3}$. The QPC features a pronounced 0.7 structure, the first plateau is not shown. (a) The blue dotted line is the derivative of the conductance $dG/dV_{\mathrm{d3}}$ (in arbitrary units). The red dashed line shows the expected contrast proportional to ${dG/dV_{\mathrm{d3}}\over G}$ of the QPC detector calculated from the pinch-off curve (black line). The black squares are direct measurements of the contrast $\Delta I_\mathrm{QPC}/I_\mathrm{QPC}$ obtained across the charging lines of QD C, where the charge changes by one electron. (b) The green dashed-dotted line is the expected shot noise spectral density of the QPC in arbitrary units. It is calculated from the pinch-off curve.}
\label{suppl_fig_1}
\end{figure}

In Suppl.\ Fig.\ \ref{suppl_fig_1}a
the relative conductance $G\,/\,G_0$ with $G_0=2e^2\,/\,h$ of QPC-I (black solid line, right axis) and its derivative $dG\,/\,dV_\mathrm{d3}$ (blue dotted line) are displayed as a function of gate voltage $V_\mathrm{d3}$. At maximum derivative the sensitivity of the QPC used as charge detector is highest. The black squares in Suppl.\ Fig.\ \ref{suppl_fig_1}a are the relative changes in the detector current $\Delta I_\mathrm{QPC}\,/\,I_\mathrm{QPC}$ measured across a charging line of QD C. It is the contrast of the detector signal. The red dashed line is calculated from the pinch-off curve (black line) with $A_0{dG/dV_{\mathrm{d3}}\over G}$, where the overall amplitude $A_{0}$ is a free  parameter. It marks the detector contrast expected from the pinch-off curve and shows a good agreement with the directly measured data (black squares). Obviously, the contrast of the detector signal is still high at already quite small conductance and small absolute detector sensitivity (blue dotted line). Suppl.\ Fig.\ \ref{suppl_fig_1}b plots the spectral density $S_\mathrm{QPC}=2eI_\mathrm{QPC}(1- {G\over{G_{\mathrm{0}}}})$ expected for the shot noise of the same QPC (green dashed-dotted line). Clearly, strong shot noise is expected at the point of highest detector sensitivity. Direct back-action related to shot noise has been studied elsewhere [S1,S2] but is not the main focus of this Letter. Instead we focus on a very different indirect back-action mechanism that often outweighs the direct back-action. Accordingly, we disclaim sensitivity and choose a working point of our detector QPC close to pinch-off, where shot noise is weak but the detector contrast is still high. The full range of working points used in the associated Letter is marked by a double arrow in Suppl.\ Fig.\ \ref{suppl_fig_1}b.

In real-time measurements [S3,S4], which need a large detector band-width, it is beneficial to use a detector working point corresponding to a high sensitivity. However, here we show that a substantial investigation of back-action is possible in low bandwidth measurements, where the detector can be used at a smaller sensitivity but less shot noise in the detector circuit.

\section{Influence of symmetry properties of the double QD on the observation of back-action}

\begin{figure}[h]
\includegraphics[width=16cm]{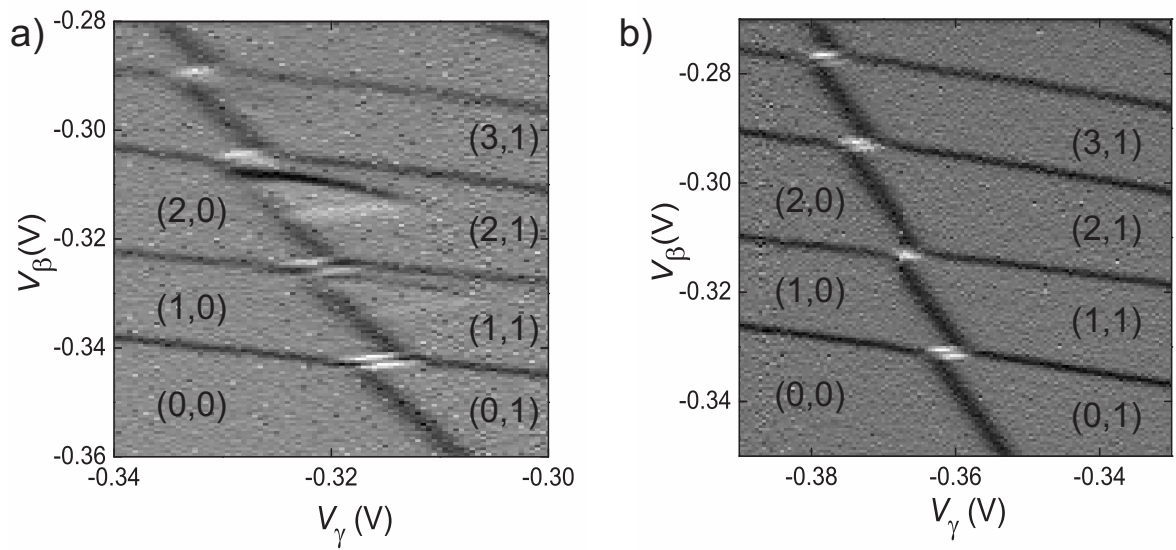}
\caption{Transconductance $dI_{\mathrm{QPC}}/dV_{\beta}$ (gray scale) as a function of $V_{\beta}$ and $V_{\gamma}$ (compare Fig.\ 1a for the geometry). QPC-I is operated at $V_{\mathrm {QPC}}=-0.8$\,mV and $P_{\mathrm {QPC}}=12$\,pW in both measurements. (a) The tunnel couplings between the two QDs and their leads are very different ($V_{\mathrm {t1}}= -245$\,mV, $V_{\mathrm {b2}}= -350$\,mV); the right tunnel barrier b2 connecting QD C with lead III is almost closed. (b) Analog measurement for more symmetric coupling to the leads ($V_{\mathrm {t1}}= -265$\,mV, $V_{\mathrm {b2}}= -330$\,mV).}
\label{suppl_fig_2}
\end{figure}

Suppl.\ Fig.\ \ref{suppl_fig_2}
plots the transconductance $dI_{\mathrm{QPC}}/dV_{\beta}$ in almost identical sections of stability diagrams for two different settings of the voltages applied to gates t1 and b2, which strongly effect the tunnel barriers between the two QDs and their respective leads (compare Fig.\ 1a of the associated Letter). The bias voltage $V_\mathrm{QPC}$ applied across QPC-I as well as the dissipated power $P_\mathrm{QPC}$ are identical for both diagrams. In Suppl.\ Fig.\ \ref{suppl_fig_2}a the barrier b2 between QD C and lead III is almost closed, resulting in a tunneling rate in the kHz-range, whereas the tunneling rate between QD B and lead II as well as the interdot tunneling rate between the two QDs are about five orders of magnitude larger. In contrast, for Suppl.\ Fig.\ \ref{suppl_fig_2}b the tunnel couplings between the two QDs and their respective leads are quite similar. In the asymmetric case shown in Suppl.\ Fig.\ \ref{suppl_fig_2}a we observe two effects caused by absorption of energy, namely the split charge reconfiguration lines (white double lines with positive slopes) and the back-action induced triangles at charge configurations (1,1), (2,1), and weakly at (3,1). In the symmetric case plotted in Suppl.\ Fig.\ \ref{suppl_fig_2}b the back-action induced triangles are missing, while some of the charge reconfiguration lines are still split.

In the following we explain why the back-action induced triangles cannot be observed for symmetric tunnel couplings between the two QDs and their leads (Suppl.\ Fig.\ \ref{suppl_fig_2}). In thermal equilibrium the occupation difference between charge configurations is given by the Boltzman distribution. At the low temperatures in our experiments a Boltzman distribution corresponds to the ground state occupation (everywhere but right at the charging and charge reconfiguration lines). Out of equilibrium deviations from a Boltzman distribution are possible. However, occupation inversion -- as observed within the triangles -- is only possible if at least three energy states are involved. For the triangles observed in Suppl.\ Fig.\ \ref{suppl_fig_2}a, the three states correspond to the configurations (1,1), (2,0) and (1,0), where (1,1) is the ground state. In case of asymmetric coupling direct transitions (1,0) $\rightarrow$ (1,1) are very slow since the tunnel barrier between QD C and its lead is almost closed (compare sketch in Fig.\ 1c). The predominant relaxation channel (1,1) $\leftrightarrow$ (2,0) $\leftrightarrow$ (1,0) involves three states, a necessary condition for occupation inversion. And we actually observe back-action induced triangles in Suppl.\ Fig.\ \ref{suppl_fig_2}a. In contrast, for equal tunnel couplings between the QDs and their respective leads the transitions (1,1) $\leftrightarrow$ (1,0) are also possible without additional occupation of an intermediate state (2,0). In particular the reoccupation of the ground state (1,0) $\rightarrow$ (1,1) can happen via one resonant tunneling process of an electron from lead III to QD C. This is the fastest tunneling process involved because it is the only one that doesn't require absorption of energy. As a direct consequence, most of the time the ground state configuration (1,1) is occupied and no back-action induced triangles can be observed (Suppl.\ Fig.\ \ref{suppl_fig_2}b).

The split charge reconfiguration lines measured in transconductance correspond to a double step of the current $I_\mathrm{QPC}$ along $\Delta$. Between the two parallel lines the current is almost constant at a value suggesting approximately equal population of the two degenerate charge states (e.\,g.\ (0,1) and (1,0) ). Here, a non-equilibrium energy source (e.g. the QPC) drives rapid oscillations between the two eigen-states, which result from the tunnel coupling between the two localized states (0,1) and (1,0). The rapid oscillations cease, if the energy splitting $E=\sqrt{\Delta_0^2+\Delta^2}$ between the eigen-states exceeds the maximum available energy $E_ \mathrm{max}$, where $\Delta_0$ is the interdot tunnel splitting and $\Delta$ the asymmetry energy between the localized states (0,1) and (1,0). In addition, the energy relaxation rate decreases with increasing $|\Delta|$ because as the eigen-states converge into the localized states (0,1) and (1,0), the overlap of the wavefunctions reduces. The distance between the split charge reconfiguration lines is determined by these two conditions. Here, we omit discussing the ratio of the interdot tunneling rate with the respective tunneling rates between the two QDs and their leads. However, in the experiments discussed here, the interdot tunnel coupling exceeds the couplings between the QDs and their leads.

Note that the charge reconfiguration line between the states (2,0) and (1,1) in Suppl.\ Fig.\ 2b is not split. Here, Pauli spin-blockade is possible [S5]. Under certain conditions spin-blockade can strongly reduce interdot tunneling. As a consequence, the otherwise observed rapid oscillations are absent and the reconfiguration line is not split. We have not observed any spin-blockade related effects on the back-action induced triangles.

\section{Strongly driven QPC-I (Phonon-mediated vs.\ Coulombic interaction)}

\begin{figure}[h]
\includegraphics[width=16cm]{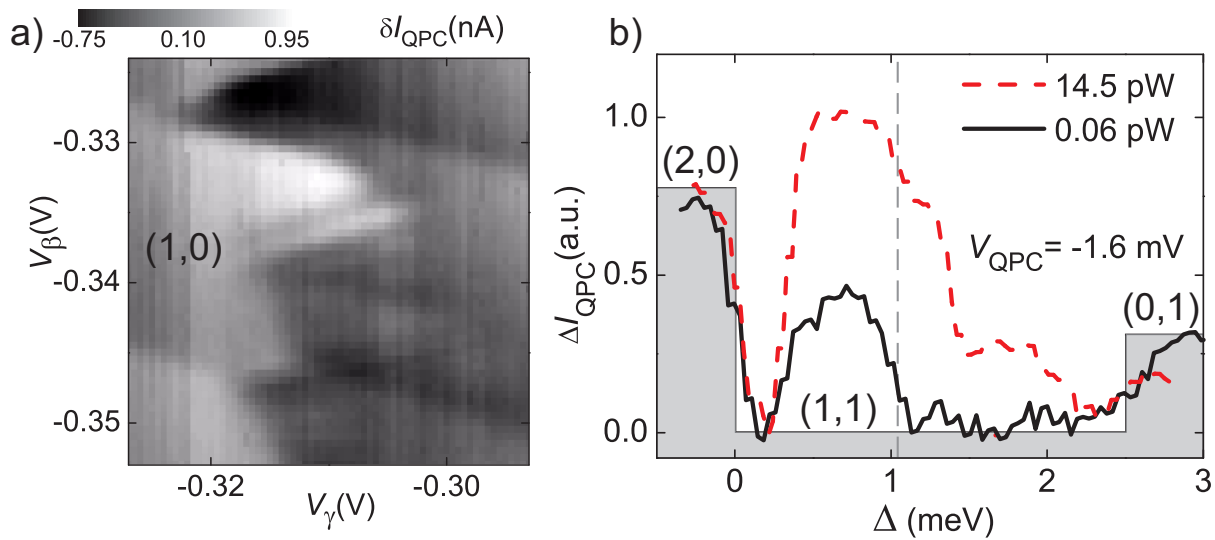}
\caption{(a) $\delta I_{\mathrm{QPC}}$ (gray scale) as a function of $V_{\beta}$ and $V_{\gamma}$. A plane fit is subtracted from $I_{\mathrm{QPC}}$. Here QPC-I is the emitter operated at $V_{\mathrm {QPC}}= -1.6$\, mV and $P_{\mathrm {QPC}}= 14.5$\,pW. (b) Normalized current change $\Delta I_{\mathrm{QPC}}$ versus asymmetry energy $\Delta$ for QPC-I at $V_{\mathrm {QPC}}= -1.6$\,mV. More details in Fig.\ 3 of the associated Letter.}
\label{suppl_fig_3}
\end{figure}

Fig.\ 3a of the associated Letter shows only one back-action induced triangle, while Fig.\ 3b contains two triangles. As described in the Letter, the upper triangle can be allocated to energy absorption in QD C while the lower triangle is related to energy absorption in QD B. From these measurements one could conclude that phonons emitted in the leads of QPC-I can only excite QD C (resulting in the upper triangle) while phonons emitted from QPC-II are absorbed by both QDs (resulting in two triangles). However, in the first case (Fig.\ 3a) QPC-I is weakly biased with $P_\mathrm{QPC}=2.5\,$pW, while for Fig.\ 3b QPC-II is driven much stronger, namely at $P_\mathrm{QPC}=72\,$pW. Suppl.\ Fig.\ \ref{suppl_fig_3}a
plots a comparable section of the stability diagram with QPC-I being strongly driven ($V_{\mathrm{QPC}} = -1.6\,$mV, $P_{\mathrm{QPC}} = 14.5\,$pW). In this case the upper triangle is clearly visible, but in addition the second (lower) triangle starts to appear. In Suppl.\ Fig.\ \ref{suppl_fig_3}b two measurements comparable to those in Fig.\ 3d of the associated Letter are shown. The power dissipated in QPC-I is $P_{\mathrm{QPC}}=0.06\,$pW for the solid line and $P_{\mathrm{QPC}}=14.5\,$pW for the red dashed line (taken from the measurement shown in Suppl.\ Fig.\ \ref{suppl_fig_3}a), while $V_{\mathrm{QPC}}=-1.6\,$mV in both cases. The smaller maximum at an asymmetry energy of $\Delta\sim1.8\,$meV belongs to the lower triangle. Nevertheless, this phonon induced maximum is superimposed by the high energy contributions of the upper triangle (also seen in Suppl.\ Fig.\ \ref{suppl_fig_3}a), which are caused by indirect Coulomb interaction and are related to the excitation spectrum of QD C, as discussed in the associated Letter.

In conclusion, phonons can be reabsorbed in both QDs no matter where they have been emitted. However, the magnitude of phonon-mediated back-action strongly depends on the detailed geometry. In our experiments the relevant phonons are emitted by relaxation of excited charge carriers in the leads of the driven QPC. For the energies typical for our measurements, the excited charge carriers usually scatter with the cold Fermi-sea before an acoustic phonon can be emitted. Energies and directions of the emitted phonons, in turn, depend on the energies and momenta of the excited charge carriers. In addition the anisotropic coupling tensors for electron-phonon interaction play a critical role [S6,S7]. In particular, phonons are emitted much stronger along certain crystal directions than other directions. While a quantitative analysis goes beyond the scope of this paper, these considerations explain, why the absorption strength of phonons in specific locations (QDs) strongly depends on the detailed geometry of emitter and detector.

For $\Delta\lesssim1.04\,$meV, where we identified phonon-mediated interaction as the main coupling mechanism, features related to the excitation spectrum of QD C are missing (e.\,g.\ Fig.\ 2a), even so excited states are expected to exist for $\Delta\lesssim1.04\,$meV. We can speculate that the absorption of phonons is accompanied with a level broadening that in turn prevents us from resolving the excitation spectrum. However, the microscopic details of this observation are not yet understood. Nevertheless, the phenomenological difference between the two discussed regimes ($\Delta\lesssim1.04\,$meV and $\Delta>1.04\,$meV) can be taken as a hint, that in our experiments back-action caused by two different coupling mechanisms is observed.

With QPC-II as emitter (Figs.\ 3b and 3e) we find only one of the two back-action mechanisms, namely the phonon-mediated back-action observed for $\Delta\lesssim1.04\,$meV. In fact, the second mechanism still seen for $\Delta>1.04\,$meV has only been observed with QPC-I as emitter and the energy being absorbed in QD C. An important difference between the two QPCs is their capacitive coupling to the QDs. The strongest capacitive coupling exists between QPC-I and QD C. In detail, a difference of one electron in QD C causes a shift of the local potential at QPC-I by approximately 5\,$\mathrm{\mu}$eV. The capacitive coupling between one of the leads of QPC-I (lead III) and QD C is expected to be even stronger. The fact that we observe the excitation spectrum of QD C only when QPC-I is strongly biased (Figs.\ 2a and 3c) indicates, that for $\Delta>1.04\,$meV Coulomb interaction is probably the predominant coupling mechanism involved. However, $E_\mathrm{max}<eV_\mathrm{QPC}$ (closed circles in Fig.\ 2c) is in contradiction to the expectation for direct Coulomb interaction between the QPC and the double QD. An indirect Coulomb interaction is possible via excited charge carriers, while they are reflected at the tunnel barrier between QD C and lead III (compare Fig.\ 1a in the associated Letter). The energy stems from ''hot'' electrons injected from QPC-I into lead III and relaxed via scattering with the Fermi-sea in lead III. The relaxation processes shift the energy spectrum compared to that directly emitted by the QPC.
\vspace{1cm}
\setlength{\parindent}{0cm}

[S1] Onac \emph{et al.,} Phys.\ Rev.\ Lett.\ {\bf 96}, 176601(2006)\newline
[S2] Gustavsson \emph{et al.,} Phys.\ Rev.\ B {\bf 78}, 035324(2008)\newline
[S3] Gustavsson \emph{et al.,} Phys.\ Rev.\ Lett. {\bf 96}, 076605 (2006)\newline
[S4] Flindt \emph{et al.,} PNAS {\bf 106}, 10116(2009)\newline
[S5] Ono \emph{et al.,} Science {\bf 297}, 1313 (2002)\newline
[S6] H. Karl \emph{et al.,} Phys.\ Rev.\ Lett.\ {\bf 61}, 2360(1988)\newline
[S7] D. Lehmann \emph{et al.,} Phys. Rev. B {\bf 65}, 085320 (2002)